\documentclass[]{edpsci}
\makeatletter
\renewcommand\@biblabel[1]{#1.}
\makeatother

\usepackage{amsmath}
\usepackage{graphicx}
\usepackage{epstopdf}
\usepackage{comment}
\usepackage[numbers,square,sort&compress]{natbib}
\usepackage{url}
\usepackage{stfloats}
\usepackage[ruled]{algorithm2e}
\usepackage[switch]{lineno}
\usepackage{hyperref}

\hypersetup{colorlinks=true,citecolor=blue,urlcolor=blue,linkcolor=blue}

\journalname{JEOS - Rapid Publication}
\articletype{Research article}

\begin{document}

\renewcommand{\inst}[1]{\unskip$^{#1}$} 

\raggedbottom
\title{Shot-to-shot statistical fluctuations during beam self-cleaning in multimode fibers}



\author{Alessia Cirimele\inst{1}
\and
Lucas Zanaglia\inst{1}
\and
Nelson Manuel  Mora Fernandez\inst{1}
\and
Alfredo Mazzulla\inst{2}\and
Pasquale Pagliusi\inst{1,2}\and
Bertrand Kibler\inst{3}\and
Guy Millot\inst{3}\and
Gabriella Cipparrone\inst{1,2}\and
Mario Ferraro\inst{1,2}\correspondingauthor{\email{mario.ferraro92@unical.it}}}
\institute{Dipartimento di Fisica, Università della Calabria, Via P. Bucci, Rende, 87036, Italy\and
CNR-NANOTEC, SS di Rende, Via P. Bucci, Rende, 87036, Italy\and
Université Bourgogne Europe, CNRS, Laboratoire Interdisciplinaire Carnot de Bourgogne ICB UMR 6303, 21000 Dijon, France}

\authorrunning{A. Cirimele et al.}

\abstract{
Beam self-cleaning in multimode optical fibers has recently been interpreted within a thermodynamic framework, in which nonlinear propagation drives the optical field towards a statistical equilibrium state. While this interpretation has been supported by measurements of modal populations and entropy, the role of shot-to-shot spatial fluctuations of the output beam has so far remained unexplored. Here, we experimentally show that the transverse output beam profile becomes progressively more reproducible from shot to shot as beam self-cleaning develops.
This behavior is consistent with the thermodynamic expectation that relative fluctuations become increasingly small as the system approaches equilibrium.}



\maketitle

\section{Introduction}

Beam self-cleaning (BSC) is one of the most remarkable nonlinear phenomena observed during light propagation in multimode optical fibers \cite{krupa2019multimode,wright2022physics}. Above a critical power threshold, an initially speckled beam spontaneously evolves into a robust bell-shaped transverse profile, despite the complex nonlinear interactions among a large number of guided modes \cite{krupa2017spatial}. Since its first observation, BSC has attracted considerable attention owing to both its fundamental implications for nonlinear wave dynamics and its potential applications in high-power fiber lasers, beam delivery, and imaging systems \cite{teugin2020single,moussa2021spatiotemporal}.

Recent theoretical advances have shown that BSC can be understood within the broader framework of optical thermodynamics \cite{ferraro2023spatial}. In this picture, the optical field behaves as a many-body system whose evolution is governed by the interplay between nonlinearity and the conserved quantities of the multimode nonlinear Schrödinger equation, notably the number of photons and the momentum of light or Hamiltonian \cite{aschieri2011condensation,wu2019thermodynamic}. The irreversible redistribution of energy among spatial modes can therefore be described as a thermalization process leading the system toward a thermodynamic equilibrium characterized by maximum entropy under the appropriate conservation laws. 

The thermodynamic interpretation of BSC has received substantial experimental support, including observations of Rayleigh--Jeans condensation \cite{baudin2020classical} and mode-resolved measurements consistent with thermalization predictions \cite{pourbeyram2022direct,mangini2022statistical}. These measurements, enabled by modal-decomposition techniques \cite{gervaziev2021mode}, also provide access to the modal entropy, whose increase and eventual saturation during the transition from a speckled to a self-cleaned beam have been directly demonstrated \cite{mangini2024maximization}. Such observations constitute the strongest experimental validations of the thermodynamic description of BSC. On this basis, we use the term 'self-cleaning regime' throughout this work to refer to conditions in which the system is expected to approach thermodynamic equilibrium.

Despite the remarkable agreement between theory and experiments, several aspects of the thermodynamic description remain under active investigation. For instance, the applicability of thermodynamic concepts to the intrinsically non-monochromatic radiation generated during nonlinear propagation is still being explored \cite{zanaglia2026spatiotemporal,genty2025full}. Generally speaking, the temporal aspects related to the BSC effect \cite{leventoux20203d,labaz2024spatial,eslami2022two} have never been considered within the thermodynamic framework.
Likewise, the coexistence of thermalization with the preservation of a high degree of spatial coherence—a distinctive feature of BSC \cite{krupa2017spatial,fabert2020coherent}—raises fundamental questions about the extent to which equilibrium statistical mechanics can describe coherent optical waves.

Among these open issues, a central one concerns the statistical basis of the thermodynamic description \cite{huang2008statistical}. While the theory relies on ensemble averaging over random-phase realizations, BSC experiments are predominantly performed using highly coherent laser sources. This raises the question of how a framework developed for incoherent waves can successfully describe experiments performed with coherent light.

In some studies, thermodynamic observables were obtained by averaging over multiple independent realizations of the optical field \cite{baudin2020classical}, whereas others successfully applied the thermodynamic framework to coherent laser beams without explicit statistical averaging \cite{pourbeyram2022direct,mangini2022statistical}. Despite the agreement of the latter approach with thermodynamic predictions, the physical origin of the statistical ensemble underlying these experiments remains largely unexplored \cite{baudin2023rayleigh}.

A plausible explanation lies in the intrinsic shot-to-shot fluctuations of pulsed laser sources. Even under nominally identical operating conditions, consecutive laser pulses inevitably exhibit small variations in energy, temporal profile, phase, and spatial distribution. These fluctuations may naturally generate an ensemble of distinct realizations, thereby providing the statistical basis required by the thermodynamic framework. This hypothesis has been suggested qualitatively in previous studies---for example, to justify ensemble averaging in BSC experiments \cite{pourbeyram2022direct,ferraro2023spatial}---but, to the best of our knowledge, no systematic experimental investigation of shot-to-shot fluctuations and their role in optical thermalization has been reported so far.

In this work, we address this gap by statistically characterizing the evolution of shot-to-shot spatial fluctuations across the self-cleaning transition in multimode optical fibers, thereby clarifying their role in defining the statistical ensemble underlying the thermodynamic description in coherent laser experiments.

The paper is organized as follows. Section 2 describes the experimental setup and the characterization of input-beam fluctuations for two laser sources: a Q-switched laser and a mode-locked laser. Section 3 presents the evolution of BSC and shot-to-shot fluctuations with input power, including the effects of spectral filtering. In Section 4, we discuss the results in the framework of optical thermodynamics, focusing on the power dependence and statistical distributions of the fluctuations, as well as the implications for modal decomposition experiments. Finally, Section 5 summarizes the main conclusions.

\begin{figure*}[!b]
\centering
\includegraphics[width=0.95\textwidth]{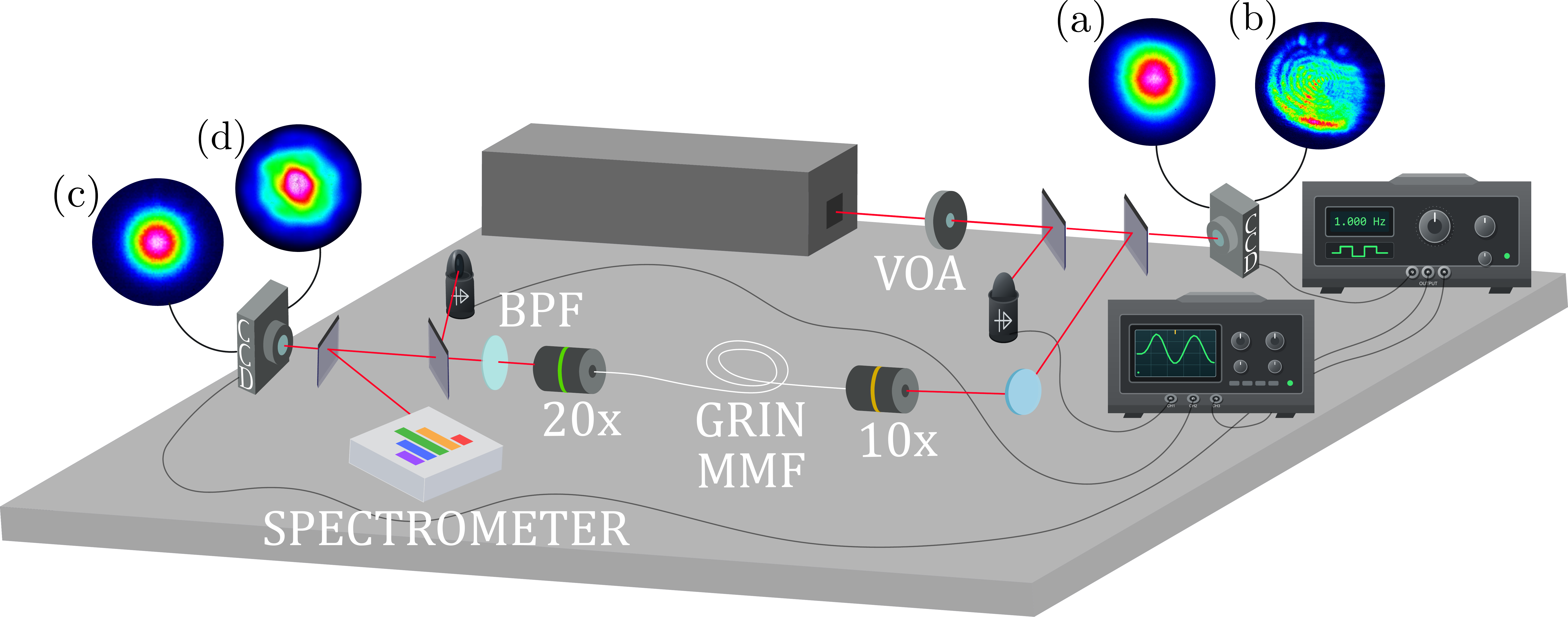}
\caption{Schematic of the experimental setup. The input beam is coupled into a 15-m-long GRIN MMF with a 20 µm core diameter using a 10X microscope objective, while the output beam is collected with a 20X objective. The input and output optical powers are simultaneously monitored by photodiode sensors and recorded with an oscilloscope synchronized with both input and output camera acquisitions. A fraction of the beam is also directed to a compact spectrometer for spectral characterization. BPF: band-pass filter; VOA: variable optical attenuator.
Insets (a–d) show experimentally acquired near-field beam profiles for the mode-locked (a, c) and Q-switched (b, d) laser sources. Panels (a, b) correspond to the input beams, whereas (c, d) show the respective output profiles in the BSC regime.
}
\label{fig-setup}
\end{figure*}

\section{Material and Methods}

BSC experiments were performed using two linearly polarized laser sources with markedly different temporal characteristics. The first source was a frequency-doubled Q-switched Nd:YAG laser (Continuum, Surelite II-10) operating at a wavelength of 532 nm, delivering 6 ns pulses at a repetition rate of 10 Hz. Whereas, the second source was a mode-locked laser (Carbide - Light Conversion) operating at a central wavelength of 1030 nm with a repetition rate of 100 kHz. The laser emitted 175 fs pulses, which were temporally stretched to approximately 30 ps by adjusting the internal pulse compressor. Pulse stretching was adopted to reduce the peak power while preserving the pulse energy, thereby allowing a direct comparison with the nanosecond experiments in a regime dominated by Kerr BSC.

In both configurations, the laser beam was coupled into a 15-m-long graded-index (GRIN) multimode fiber with a 20 µm core diameter using a 10X microscope objective. The input power was controlled by means of a variable attenuator. The fiber supports approximately 40 guided spatial modes at 532 nm and 10 modes at 1030 nm \cite{stefanska2024high}. Similarly, a 20X microscope objective was used to collect the beam at the fiber output. A sketch of the experimental set-up is illustrated in Fig. \ref{fig-setup}.

The input and output near-field energy distributions were recorded using two separate beam-profiling cameras (Beamage-4M, Gentec, with a refresh time of 1s). We fixed the exposure time to its minimum value of 0.1 ms in all the experiments. In this way, owing to the low repetition rate of the Q-switched laser, the near-field profile of each individual pulse could be directly recorded, enabling a true single-pulse statistical analysis; whereas, in the case of the mode-locked laser, each camera frame corresponded to a train of 10 consecutive pulses. To ensure that the CCD camera operated in a fully linear response regime, neutral-density filters were placed in front of the camera so that the recorded energy levels remained approximately within the same grayscale range.

\subsection{Q-switched laser}
In the case of the Q-switched laser, the low repetition rate prevented us from directly measuring the energy of each individual pulse using a photodiode. We therefore calculated the shot-to-shot energy fluctuations from the single-pulse images acquired with the CCD camera. The analysis was based on a set of 100 images, one per laser pulse, indexed by $k$, with the $k$-th image denoted by $I_k^{in}$.

For each pulse, we first calculated the integrated CCD signal as 
\begin{equation}
S_k^{in} = \sum_{i,j} I_k^{in}(i,j),
\end{equation}
where the sum extends over all pixels of the $k$-th image. We then defined the normalized quantity
\begin{equation}
N_k^{in} =
\frac{S_k^{in}}
{\displaystyle\max_{k}\left(S_k^{in}\right)},
\label{eq:Nk}
\end{equation}
where the maximum is taken over the 100 acquired pulses. Thus, $N_k^{in}$ provides a measure of the relative shot-to-shot fluctuations of the input energy 
and can be regarded as a proxy for the fluctuations of the photon number ($N$) emitted by the laser in each pulse.

\begin{figure}[!ht]
\centering
\includegraphics[width=\columnwidth]{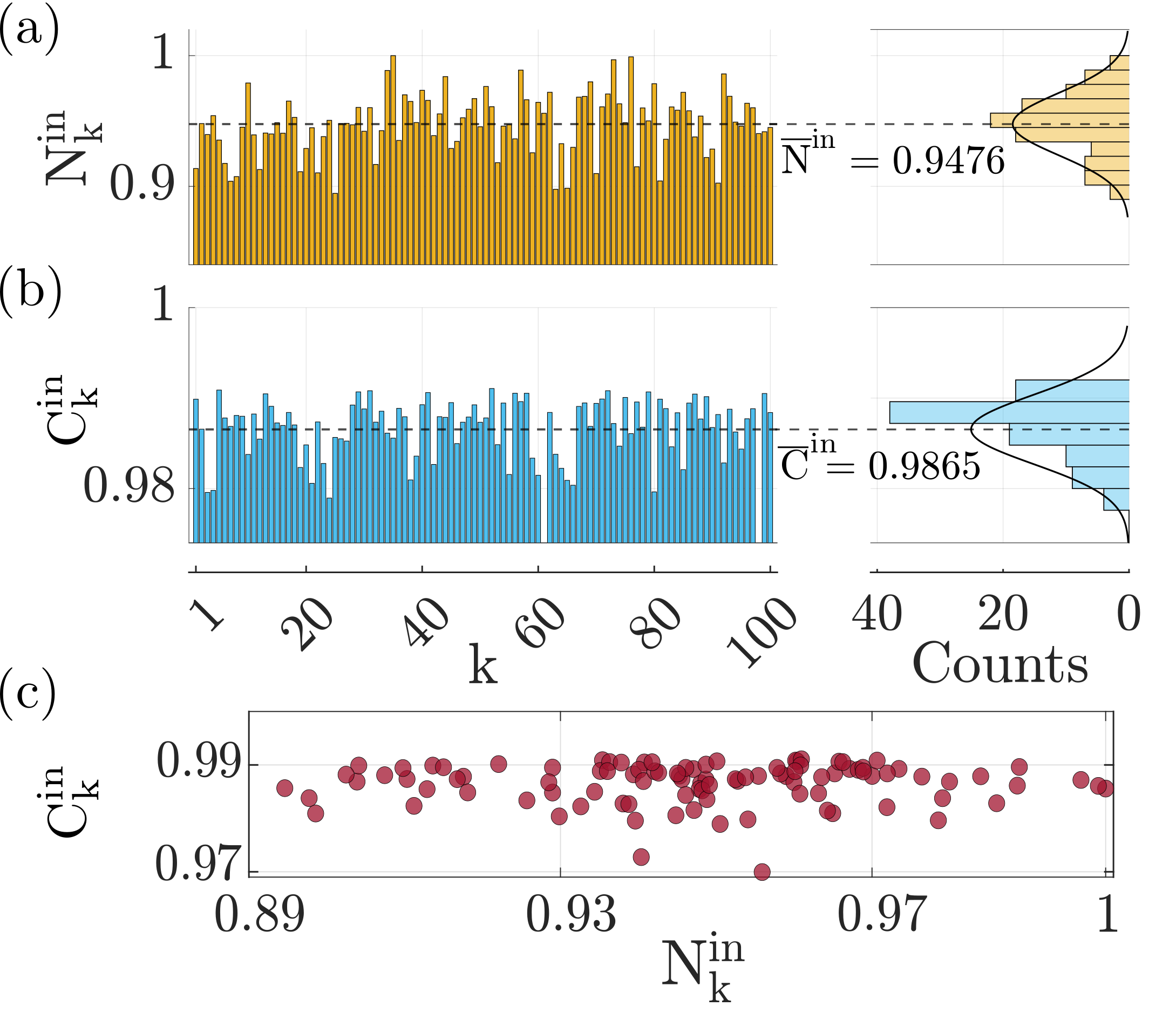}
\caption{Statistical characterization of the input beam for the Q-switched laser.
(a) Normalized energy $N_k^{\mathrm{in}}$ for the 100 consecutive laser pulses, together with its corresponding distribution. The dashed line indicates the mean value.
(b) Correlation coefficient $C_k^{\mathrm{in}}$ between each single-shot input-beam profile and the ensemble-averaged profile, shown as a function of the pulse index $k$. The dashed line indicates the mean correlation, $\overline{C}^{\mathrm{in}} = 0.9865$, and the right panel shows the corresponding distribution.
(c) Correlation coefficient $C_k^{\mathrm{in}}$ as a function of $N_k^{\mathrm{in}}$, showing no evident dependence of the spatial profile on the shot-to-shot energy fluctuations.}\label{fig2-sorgenteQ}
\end{figure}

The resulting distribution is shown in Fig. \ref{fig2-sorgenteQ}a. As can be seen, $N_k^{in}$ exhibits appreciable shot-to-shot fluctuations, with the measured values approximately following a bell-shaped distribution. The distribution has a mean value of 0.948 and a standard deviation of 0.024. Despite the observed fluctuations, the relatively small width of the distribution indicates that the total photon number can be considered approximately constant across the different realizations. This represents one of the fundamental requirements underlying the thermodynamic description of the optical field.

A second quantity that is expected to remain approximately conserved within the thermodynamic framework is the Hamiltonian ($H$). Although our measurements do not provide direct access to the full Hamiltonian, changes in the input-beam spatial profile would modify the initial modal populations and, consequently, the thermodynamic parameters associated with the optical field, such as its temperature and chemical potential.
We therefore used the experimentally accessible spatial properties of the input beam to assess the reproducibility of the initial conditions across different realizations.
To this end, we quantified the similarity between each single-shot input image $I_k^{in}$ and the ensemble-averaged input energy distribution, defined as
\begin{equation}
\overline{I}^{in} =
\frac{1}{100}
\sum_{k=1}^{100} I_k^{in}.
\label{eq:average}
\end{equation}
Generally speaking, in the following, we will use the bar on top of a given symbol to define the average over 100 images as in Eq. (\ref{eq:average}).
For each realization, we calculated the correlation coefficient
\begin{equation}
C_k^{in}=
\frac{
\left[
\displaystyle\sum_{i,j} I_k^{in}(i,j)\,\overline{I}^{in}(i,j)
\right]^2
}{
\displaystyle
\sum_{i,j} \left[I_k^{in}(i,j)\right]^2 \, 
\sum_{i,j} \left[\overline{I}^{in}(i,j)\right]^2 
}.
\label{eq:C_k^in}
\end{equation}

As shown in Fig.~\ref{fig2-sorgenteQ}b, $C_k^{in}$ remains close to unity across all 100 pulses, indicating a highly reproducible input-beam profile and suggesting limited shot-to-shot variations in the initial modal populations.

Fig.~\ref{fig2-sorgenteQ}c shows $C_k^{in}$ as a function of $N_k^{in}$. The data form an approximately horizontal cloud of points, with no evident systematic dependence of the spatial correlation on the pulse energy. This indicates that fluctuations in the total photon number are not systematically accompanied by changes in the spatial distribution of the input beam. From a thermodynamic perspective, these results suggest that the different pulses provide distinct realizations of the optical field characterized by approximately stable initial modal populations.

Finally, it is worth noting that, in BSC experiments with the Q-switched laser, the input power was varied by using a set of neutral-density filters whose attenuation was independently characterized. The input power was therefore estimated from the laser power measured upstream of the filters using a thermal power sensor, together with the calibrated attenuation of the filter set. Accordingly, the input powers reported in the following section are expressed in watts.

\subsection{Mode-locked laser}

Unlike the Q-switched case, 
the input and output optical powers were measured using two photodiode power sensors (Thorlabs S120C and S121C, respectively), connected via BNC cables to an oscilloscope (Siglent SDS5034X). The oscilloscope and the two cameras were synchronized using a common square-wave signal generated by a signal generator (Siglent SDG6022X). The signal had an amplitude of 1 V, a period of 1 s, and a duty cycle of 0.01\%. The period was chosen to match the refresh time of the cameras, while the ON time was set to match their exposure time. The same signal was used to trigger the camera acquisitions and to gate the oscilloscope traces recorded from the photodiode sensors. In this way, each camera acquisition was associated with the corresponding power measurement of the same pulse train (see Appendix \ref{app-oscilloscope}).

A 100-s-long acquisition was performed, yielding a set of 100 recorded pulse trains.  We denote by $a_k^{in}$, with $k=1,2,\ldots,100$, the average amplitude of the $k$-th pulse train. Then, we normalized the $\{a_k^{in}\}$ by their maximum, defining the quantity $\{A_k^{in}\}$ in analogy to Eq. (\ref{eq:Nk}). The latter provides a measure of the shot-to-shot fluctuations of the optical signal and is therefore proportional to the energy associated with each camera acquisition. It thus plays a role analogous to $N_k^{in}$ in the characterization of the Q-switched laser, providing a proxy for the fluctuations of the photon number associated with each realization. It should be noted that averaging over 10 consecutive pulses may partially suppress the intrinsic pulse-to-pulse fluctuations, depending on the degree of correlation between successive pulses, which was not independently measured.

\begin{figure}[!ht]
\centering
\includegraphics[width=\columnwidth]{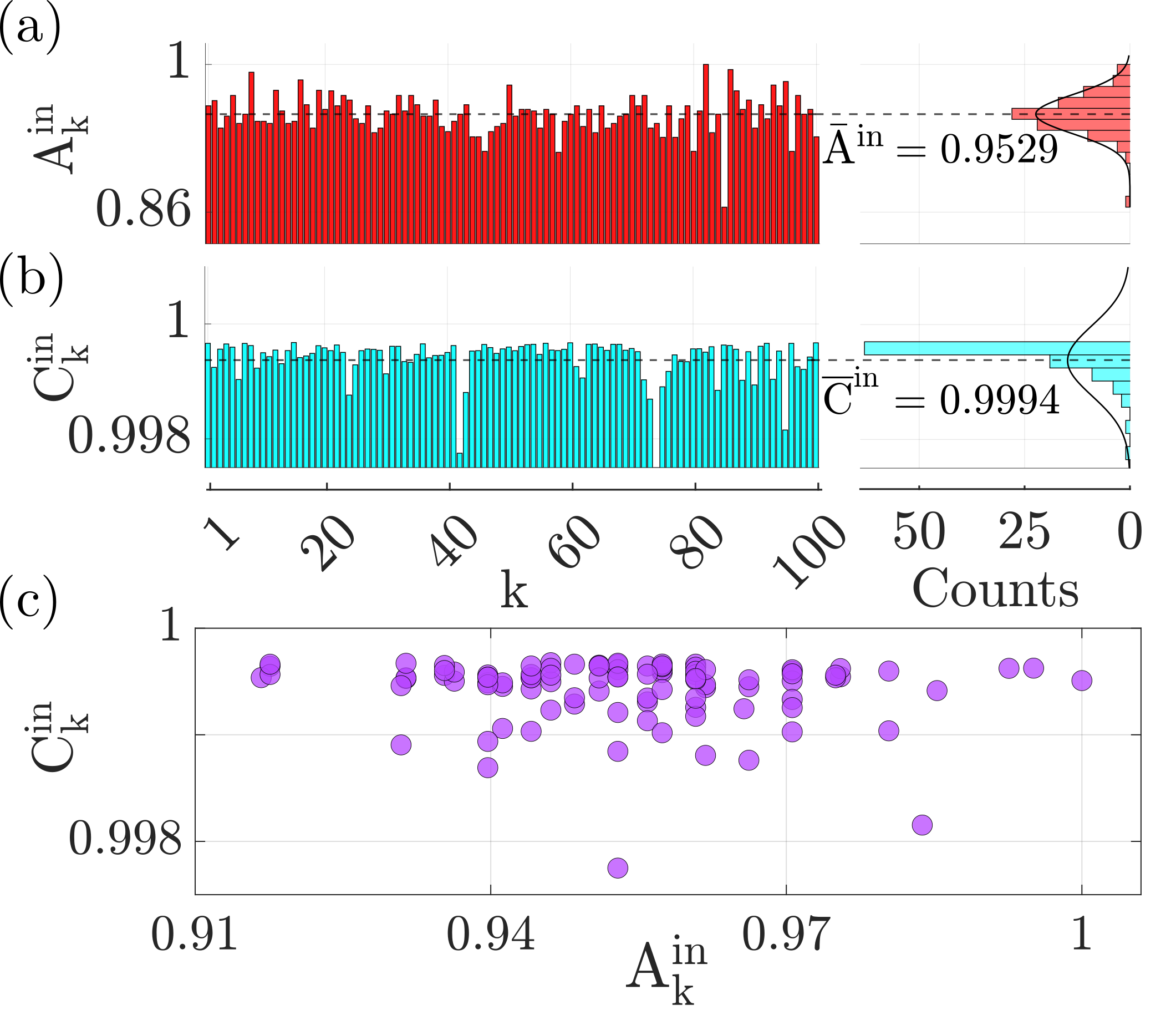}
\caption{Statistical characterization of the input beam for the mode-locked laser.
(a) Normalized average amplitude $A_k^{\mathrm{in}}$ of the 10-pulse train associated with each camera acquisition 
for 100 consecutive realizations, together with its corresponding distribution. The dashed line indicates the mean value.
(b) Correlation coefficient $C_k^{\mathrm{in}}$ between each single-shot input-beam profile and the ensemble-averaged profile, shown as a function of the realization index $k$. The dashed line indicates the mean correlation, $\overline{C}^{\mathrm{in}} = 0.9994$, and the right panel shows the corresponding distribution.
(c) Correlation coefficient $C_k^{\mathrm{in}}$ as a function of $A_k^{\mathrm{in}}$, showing no evident dependence of the spatial profile on the shot-to-shot fluctuations of the measured optical signal.} 
\label{fig3-sorgenteML}
\end{figure}

As shown in Fig. \ref{fig3-sorgenteML}a, $A_k^{in}$ fluctuates from one shot to another, with the measured values following a bell-shaped distribution. The distribution is characterized by a mean value of 0.953 and a standard deviation of 0.019. Despite these fluctuations, the relatively narrow distribution indicates that the photon number remains approximately constant across the different shots.

As in the case of the Q-switched laser, the spatial correlation coefficient $C_k^{in}$ remains close to unity for all shots (Fig. \ref{fig3-sorgenteML}b). In the case of the mode-locked laser, the spatial profile of the input beam is even more stable, with an average correlation coefficient exceeding 0.999. Furthermore, the experimental fluctuations of $A_k^{in}$ appear to be uncorrelated with those of $C_k^{in}$, as shown in Fig. \ref{fig3-sorgenteML}c. This indicates that the shot-to-shot fluctuations generate different realizations of the optical field without significantly modifying its spatial structure, consistently with the existence of a statistical ensemble characterized by approximately identical thermodynamic constraints.

Before presenting the experimental results on BSC, we briefly discuss the spectral characterization of the mode-locked laser. The pulses entering the fiber had a spectral bandwidth of approximately 20~nm, which remained essentially unchanged after propagation over the entire investigated power range (see Appendix \ref{app-spectra}). The input and output spectra were measured using a compact spectrometer (Ocean Optics FLAME-T-XR1-ES). The spectral resolution of this instrument was not sufficient to resolve the spectrum of the Q-switched laser pulses.

To investigate the dependence of BSC on the spectral bandwidth, the output beam of the mode-locked laser was also spectrally filtered using a set of band-pass filters centered at 1030~nm and characterized by different spectral bandwidths. 

\subsection{Correlation-based characterization of the output beam}
\label{subsec:output_correlations}

To characterize BSC and its statistical fluctuations, we adopted a correlation-based analysis of the output-beam. Holographic modal decomposition, commonly employed in thermodynamic experiments, would in principle provide direct access to the modal populations. However, such measurements typically require relatively long acquisition times and are therefore not compatible with the single-shot, or single-shot-like, analysis considered here.

We therefore quantified the similarity between each output energy distribution $I_k^{out}$ and and the profile of the fundamental mode $I_0$, which has a gaussian shape, through the correlation coefficient
\begin{equation}
n_{0,k}=
\frac{
\left[
\displaystyle\sum_{i,j} I_k^{out}(i,j)\,I_0(i,j)
\right]^2
}{
\displaystyle
\sum_{i,j} \left[I_k^{out}(i,j)\right]^2
\,\sum_{i,j} \left[I_0(i,j)\right]^2
}.
\label{eq:n0_corr}
\end{equation}

The notation $n_{0,k}$ is chosen to recall the conventional notation for the occupation of the fundamental mode in the thermodynamic framework ($n_0$). We emphasize, however, that $n_{0,k}$ does not represent a direct measurement of the modal occupation, but rather a correlation-based proxy for the degree of fundamental-mode content of the output beam.

To quantify the stability of the output beam, we also calculated the correlation coefficient $C_k^{out}$ between each output energy distribution $I_k^{out}$ and the ensemble-averaged output profile $\overline{I}^{out}$, following the same procedure adopted for the input-beam correlation (Eq. \ref{eq:C_k^in}). Values of $C_k^{out}$ close to unity indicate highly reproducible output profiles, whereas lower values correspond to stronger spatial fluctuations between different realizations.

Finally, in order to gain a comprehensive understanding of the evolution of the BSC and its statistical fluctuations, we calculated two complementary observables: the average fundamental-mode correlation $\overline{n}_0$, as measure of the degree of BSC, and the average output-beam correlation $\overline{C}^{\mathrm{out}}$. A high value of $\overline{C}^{\mathrm{out}}$ indicates that the individual output images are highly similar to one another, corresponding to weak shot-to-shot fluctuations.

\section{Results}
We now examine the dependence of $\overline{n}_0$ and $\overline{C}^{\mathrm{out}}$ on power for both laser sources, focusing on the relationship between BSC and the reproducibility of the output beam.

\subsection{Q-switched laser}

\begin{figure}[!htb]
\centering
\includegraphics[width=\columnwidth]{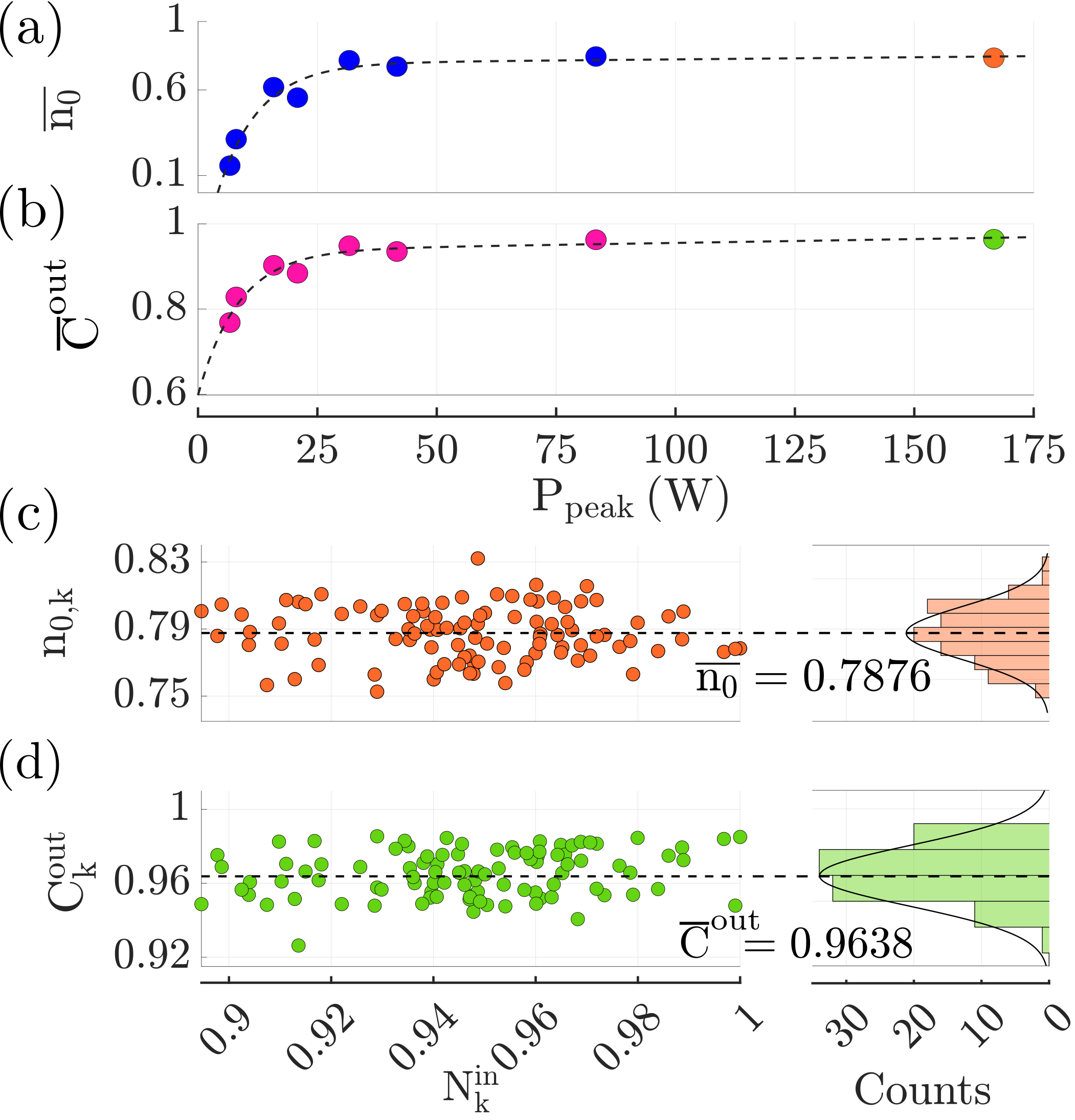}
\caption{Beam self-cleaning and statistical fluctuations for the Q-switched laser.
(a) Average fundamental-mode correlation $\overline{n}_0$ as a function of the input power, quantifying the degree of BSC. 
(b) Average output-beam correlation $\overline{C}^{\mathrm{out}}$ as a function of the input power, quantifying the reproducibility of the output beam across different realizations. 
(c) Distribution of the fundamental-mode correlation $n_{0,k}$, corresponding to the orange marker in panel (a).
(d) Distribution of the output-beam correlation $C_k^{\mathrm{out}}$, corresponding to the green marker in panel (b).}\label{fig4-corrQ}
\end{figure}

The results obtained with the Q-switched laser are reported in Fig. \ref{fig4-corrQ}. Figure \ref{fig4-corrQ}a clearly shows the onset of BSC as the input power is increased. In particular, $\overline{n}_0$ increases significantly with the input power, from approximately 0.16 at the lowest peak power ($7$ W) to 0.79 at the highest peak power ($0.17$ kW). This increase is the fingerprint of the occurrence of BSC.

At the same time, the statistical fluctuations between consecutive realizations decrease as the input power grows larger, as indicated by the enhancement of $\overline{C}^{\mathrm{out}}$ shown in Fig. \ref{fig4-corrQ}b. Thus, the occurrence of BSC is accompanied by an increasing reproducibility of the output beam from pulse to pulse. The shot-to-shot distributions of $n_{0,k}$ and $C_k^{\mathrm{out}}$ at the highest input power are shown in Figs. \ref{fig4-corrQ}c and \ref{fig4-corrQ}d, respectively. These data correspond to the orange and green points in Figs. \ref{fig4-corrQ}a and \ref{fig4-corrQ}b, respectively. As in the characterization of the input beam reported in Fig. \ref{fig2-sorgenteQ}, the values of $n_{0,k}$ exhibit an approximately bell-shaped distribution. In contrast, the distribution of $C_k^{\mathrm{out}}$ is strongly concentrated towards unity, with an average value of 0.9638. This high correlation demonstrates that, despite the shot-to-shot fluctuations of the input laser, the output beam becomes highly reproducible in the self-cleaning regime.

To exclude the possibility that the increase of $\overline{C}^{\mathrm{out}}$ with input power observed in Fig. \ref{fig4-corrQ}b results from a power-dependent sensitivity of the CCD camera, rather than from a genuine reduction of the statistical fluctuations of the output beam, we performed additional measurements. We modified the camera signal level independently of the input power. This allowed us to assess the possible influence of the signal-to-noise ratio on the calculated correlations. The results of this experiment are reported in Appendix \ref{app-noise}, where we show that the observed increase of $\overline{C}^{\mathrm{out}}$ cannot be attributed to changes in the camera signal-to-noise ratio.

\subsection{Mode-locked laser}

\begin{figure}[!ht]
\centering
\includegraphics[width=0.95\columnwidth]{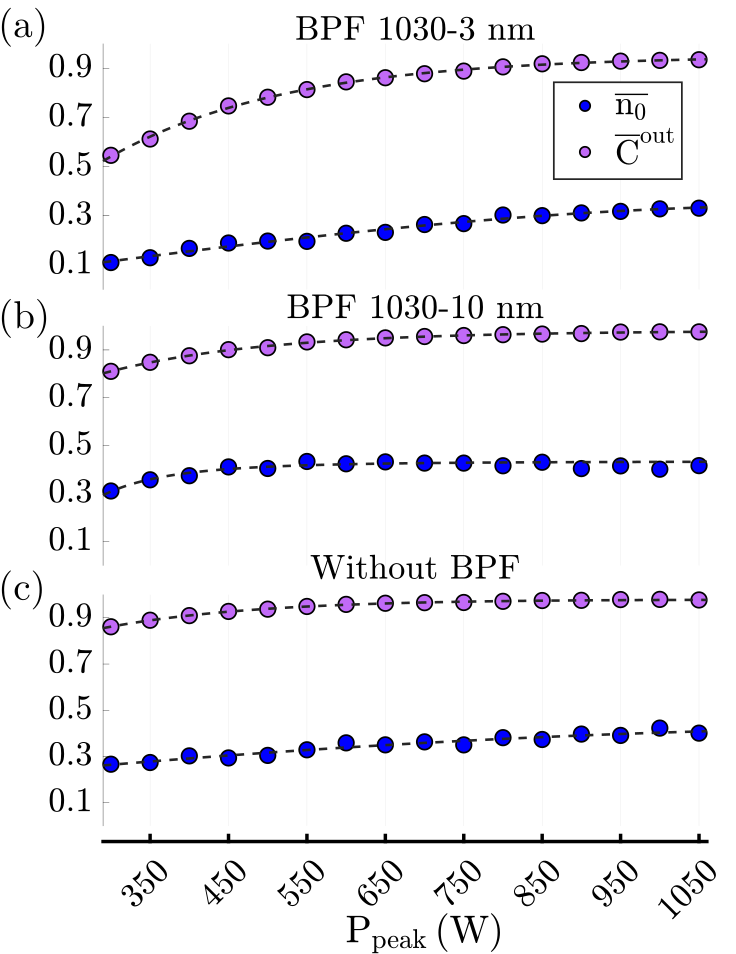}
\caption{Beam self-cleaning and statistical fluctuations for the mode-locked laser: average fundamental-mode correlation $\overline{n}_0$ and average output-beam correlation $\overline{C}^{\mathrm{out}}$ as functions of input power. Measurements were performed with band-pass filters of (a) 3~nm and (b) 10~nm bandwidth, placed at the fiber output, and (c) without spectral filtering. For each input-power setting, the three configurations were measured sequentially while keeping the input power and injection conditions unchanged.}
\label{fig5-corrML}
\end{figure}

The results obtained with the mode-locked laser (Fig. \ref{fig5-corrML}) are qualitatively consistent with those observed for the Q-switched source. In particular, BSC, manifested by the increase of $\overline{n}_0$ with input power, is accompanied by an increase of $\overline{C}^{\mathrm{out}}$, indicating a progressive reduction of the shot-to-shot fluctuations of the output beam.

To investigate the influence of spectral filtering, we acquired three consecutive sets of output-beam images at each fixed input power: first without spectral filtering, and then with band-pass filters of 3 and 10~nm bandwidth, respectively, placed at the fiber output. Throughout these acquisitions, the input power and injection conditions were kept unchanged, so that the three measurements corresponded to the same nominal propagation conditions.

For all three configurations, both $\overline{n}_0$ and $\overline{C}^{\mathrm{out}}$ increase monotonically with input power, indicating that the qualitative behavior of the BSC process is preserved across the different spectral selections. In particular, $\overline{C}^{\mathrm{out}}$ progressively approaches unity in all cases, showing that the reproducibility of the output beam is not significantly affected by the spectral filtering. In contrast, the maximum value reached by $\overline{n}_0$ depends on the selected spectral bandwidth. For the narrowest, 3-nm-wide filter, $\overline{n}_0$ reaches approximately 30\% (Fig. \ref{fig5-corrML}a), whereas values of about 40\% are obtained with the 10-nm-wide filter and without spectral filtering (Figs. \ref{fig5-corrML}b and \ref{fig5-corrML}c, respectively). These results are consistent with previous observations by Genty et al., who showed that the 'cleanliness' of the beam at high power is progressively lost as the filtered spectral bandwidth narrows \cite{genty2025full}. Nevertheless, for bandwidth as large as tens of nanometers, the bell shape of the beam turned out to be preserved.

\section{Discussion}

The central question addressed in this work is whether the shot-to-shot fluctuations observed in BSC are compatible with a thermodynamic description of the nonlinear propagation. So far, we have shown that the increase of the fundamental-mode content with input power is accompanied by a progressive reduction of the statistical fluctuations of the output beam. Now, we discuss this behavior in the framework of statistical thermodynamics and its implications for the interpretation of BSC.

\subsection{Power dependence of the fluctuations}

One of the characteristic properties of statistical thermodynamics is the smallness of relative fluctuations at equilibrium. For an extensive observable, under the assumption of sufficiently short-ranged correlations, the relative standard deviation scales as the inverse square root of the number of particles, i.e., $\sigma_X/\overline{X} \propto 1/\sqrt{N}$, so that fluctuations become progressively less important as the number of particles increases \cite{huang2008statistical}. This inverse-square-root dependence provides a reference for investigating fluctuation scaling in the optical system considered here.

We therefore investigated the dependence of the relative fluctuations of the fundamental-mode content, $\sigma_{n_0}/\overline{n}_0$, on the input power. The analysis was restricted to selected high-power points in the plateau regions of $\overline{n}_0$, associated with the self-cleaning regime and interpreted as the conditions closest to thermal equilibrium. Specifically, we considered only the six highest-power data points from Fig.~\ref{fig4-corrQ}(a) and eight highest-power blue circles in Fig.~\ref{fig5-corrML}(b), corresponding to the Q-switched and mode-locked sources, respectively. A power-law dependence was empirically introduced in the form
\begin{equation}
    \frac{\sigma_{n_0}}{\overline{n}_0} \propto P^{-\alpha},
\end{equation}
where $\alpha$ quantifies the rate at which the relative fluctuations decrease with increasing input power $P$.
In the simplest thermodynamic picture, one would expect $\alpha=1/2$ if the input power could be identified directly with the number of photons, since $N\propto P$. However, in the present experiment the input power plays a dual role. On the one hand, it determines the photon number and therefore the
magnitude of the equilibrium fluctuations. On the other
hand, it controls the strength of the Kerr nonlinearity
and hence the rate at which the system approaches its
thermodynamic equilibrium.

\begin{figure}[!htb]
\centering
\includegraphics[width=\columnwidth]{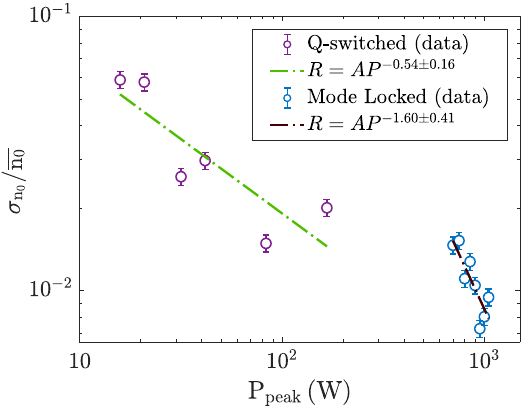}
\caption{Power dependence of the relative fluctuations of the fundamental-mode content, close to thermal equilibrium.
Relative standard deviation $\sigma_{n_0}/\overline{n}_0$ as a function of the input peak power $P_{peak}$ for the mode-locked (blue circles) and Q-switched (purple circles) laser sources. Error bars represent the estimated standard errors, assuming approximately Gaussian statistics and independent realizations. The dash-dotted lines show power-law fits of the form $R=AP^{-\alpha}$, obtained through unweighted linear regression in log-log space, yielding $\alpha=1.60\pm0.41$ for the mode-locked source and $\alpha=0.54\pm0.16$ for the Q-switched source. The reported uncertainties on $\alpha$ correspond to the standard errors of the regression slopes.}
\label{fig6-fluct_Q_ML}
\end{figure}

This dual role of $P$ provides a framework for interpreting the scaling behaviors observed with the two laser sources (Fig.~\ref{fig6-fluct_Q_ML}). The Q-switched data, obtained at peak powers on the order of $100$~W, yield a fitted exponent of $\alpha=0.54\pm0.16$, close to the reference value of $1/2$. By contrast, the mode-locked measurements, performed at peak powers approximately one order of magnitude higher, yield a substantially larger exponent, $\alpha=1.60\pm0.41$.
This wider departure from the theoretical value may be associated with the greater relative contribution of the nonlinear term to the Hamiltonian at higher peak powers. The stronger Kerr interaction can accelerate the approach to equilibrium. Consequently, the reduction of fluctuations observed over a fixed fiber length is faster than that expected from the increase of the photon number alone.

\subsection{Statistics of the fluctuations}
A further aspect concerns the statistical distribution of the shot-to-shot fluctuations. As discussed above, the distribution of $n_{0,k}$ obtained from the standard dataset does not exhibit an immediately recognizable Gaussian profile (Fig. \ref{fig4-corrQ}c). This observation is relevant in the context of the thermodynamic description of BSC, where random linear mode coupling and the resulting statistical correlations between fiber modes play a central role in the thermalization process \cite{makris2020statistical,mangini2024maximization}. Since the apparent shape of the distribution may be affected by finite statistical sampling, we performed an additional analysis using a larger statistical ensemble. The corresponding results are reported in Appendix \ref{app-1000acq}.

\subsection{Implications for modal decomposition experiments}

Our results also provide relevant insights into the robustness of the statistical properties of BSC against spectral filtering. Modal decomposition measurements commonly used to investigate the thermodynamic nature of BSC often rely on spatial light modulators, which operate over a limited spectral bandwidth. Consequently, a spectral filtering stage is typically required to select a sufficiently narrow spectral component of the output beam. Since the nonlinear propagation of ultrashort pulses is accompanied by spectral broadening, it is important to assess whether the spectral bandwidth selected at the fiber output influences the statistical properties of the self-cleaning process.

The results obtained with the mode-locked laser show that the progressive increase of $\overline{C}^{\mathrm{out}}$ towards unity is remarkably robust with respect to the spectral bandwidth selected at the fiber output. In particular, using the 10-nm-wide filter leads to $\overline{C}^{\mathrm{out}}$ values very similar to those obtained without applying filters. 
These findings suggest that the statistical reduction of shot-to-shot fluctuations accompanying BSC remains observable after spectral filtering, at least within the bandwidth range investigated here.

Our outcomes are particularly relevant to experiments requiring spectral filtering for optical characterization. However, the persistence of the observed correlation trends alone does not prove that all quantities inferred from spectrally filtered measurements, including modal populations, are independent of the selected bandwidth. Further investigation is needed to determine the extent to which such quantities are affected by spectral selection.

\section{Conclusions}

We have experimentally characterized the shot-to-shot fluctuations of the input and output beams during BSC, using both a Q-switched and a mode-locked laser source.

We have shown that the input field exhibits reproducible but non-negligible fluctuations from one realization to another. Importantly, such fluctuations are also observed for the mode-locked laser, despite the fact that the pulses generated by such sources are commonly regarded as nearly identical. The presence of these fluctuations provides a 
possible experimental source of statistical realizations in the nonlinear propagation process. In a Kerr medium, fluctuations of the optical power are associated with fluctuations of the nonlinear phase acquired through self-phase modulation, and can therefore contribute to the different phase configurations underlying the statistical evolution of the multimode field.

Most importantly, we have shown that the shot-to-shot fluctuations of the output beam are progressively reduced as BSC develops. The increase of the fundamental-mode content with input power is systematically accompanied by an increase of the correlation between different output realizations, demonstrating that the self-cleaned beam becomes progressively more reproducible from shot to shot. This behavior is in agreement with the thermodynamic picture, according to which relative fluctuations become small when a large number of particles participate in an equilibrium state.

Our results therefore provide a new experimental signature supporting the thermodynamic interpretation of BSC: beyond the observation of modal thermalization, we demonstrated that, as thermalization proceeds, the statistical fluctuations of the optical field themselves evolve towards a more stable and reproducible state. 

\acknowtext

We thank Massimino Sposato, Giovanna Palermo, Dante Maria Aceti, and Attilio Golemme for fruitful collaboration.

\funding
We acknowledge the support from the Italian Ministry of University and Research (MUR) through the Italian Science Fund (FIS2), project SOFT (H53C24001530001).

\conflict
The authors have nothing to disclose.

\dataavailability
The data that support the findings of this study are available from the corresponding author upon reasonable request.

\authorcontrib
Conceptualization, A.C. and M.F.; Methodology, A.M., P.P.; Validation, B.K., G.M. and G.C.; Formal Analysis, A.C.; Investigation, A.C., N.M.M.F.; Resources, G.M., G.C., M.F.; Data Curation, A.C.; Writing – Original Draft Preparation, A.C., M.F.; Writing – Review \& Editing, L.Z., A.M., P.P., B.K., G.M., G.C.; Supervision, Project Administration, and Funding Acquisition, M.F..

\bibliographystyle{jeos}
\bibliography{biblio_jeos}

\appendix

\section{Oscilloscope trace}
\label{app-oscilloscope}

\begin{figure}[!htb]
    \centering
    \includegraphics[width=\columnwidth]{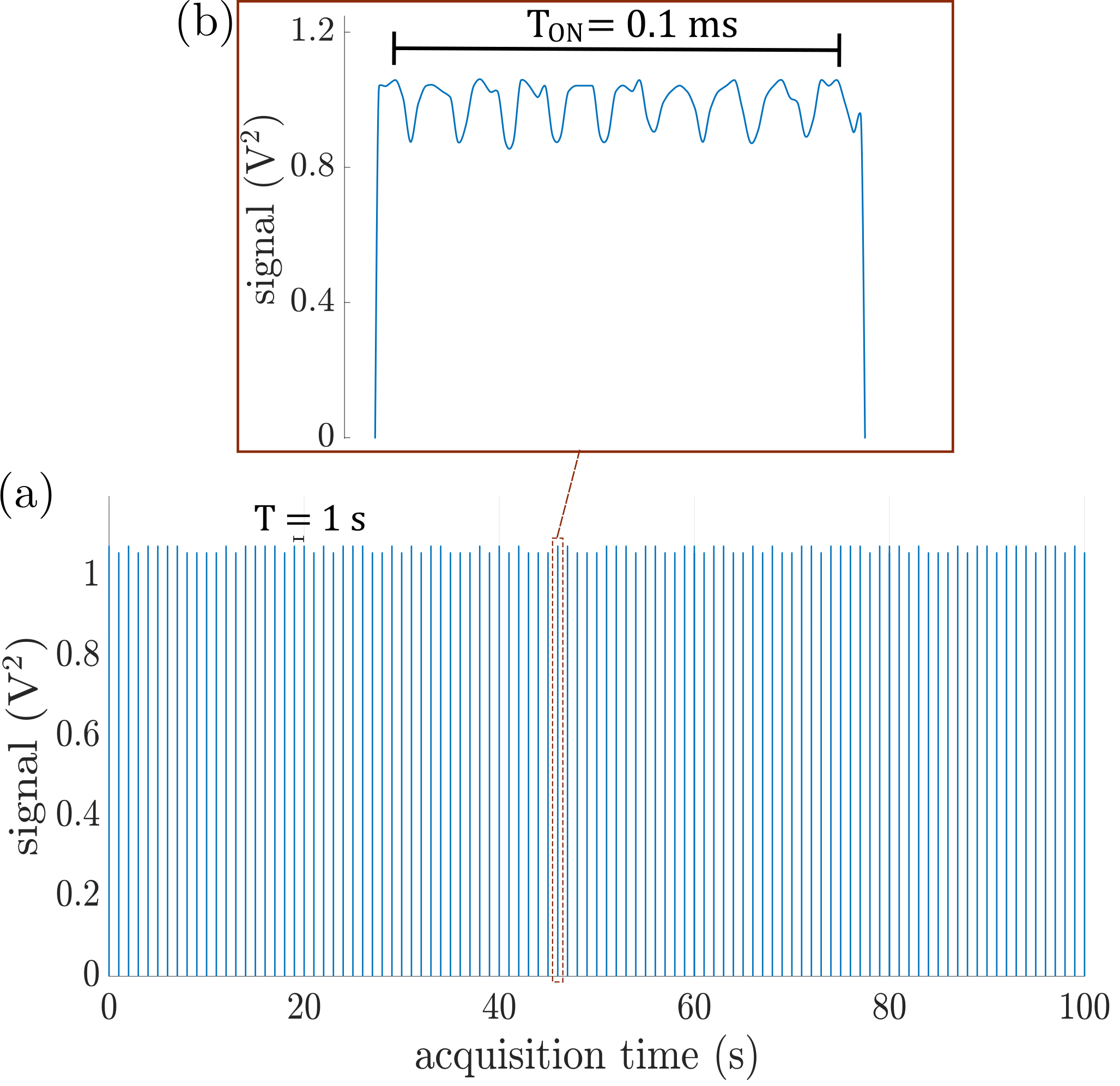}
    \caption{Structure of the acquired input signal. (a) Signal trace acquired over 100 s, consisting of 100 pulse trains repeated with a period \(T = 1\,\mathrm{s}\). (b) Magnified view of a single pulse train, showing the 10 constituent pulses within a duration of $0.1\,\mathrm{ms}$.}
    \label{fig:trace}
\end{figure}

Fig.~\ref{fig:trace} illustrates the temporal structure of the input signal acquired by the oscilloscope. The acquisition is controlled by the synchronization signal generated by the signal generator, which has a period $T=1\,\mathrm{s}$ and an ON time $T_{\mathrm{ON}}=0.1\,\mathrm{ms}$. As shown in Fig.~\ref{fig:trace}a, the resulting input signal consists of a sequence of pulse trains repeated every $1\,\mathrm{s}$. Over an acquisition time of $100\,\mathrm{s}$, 100 pulse trains are therefore recorded.

A magnified view of one acquisition window is shown in Fig.~\ref{fig:trace}b. During the $T_{\mathrm{ON}}=0.1\,\mathrm{ms}$ interval, the signal contains 10 consecutive laser pulses. The corresponding pulse repetition period is therefore approximately $10\,\mu\mathrm{s}$, corresponding to a repetition rate of $100\,\mathrm{kHz}$. The synchronization window thus defines the temporal interval over which the oscilloscope records the photodiode signal and during which the cameras acquire the corresponding optical image.

This acquisition scheme provides a direct temporal correspondence between the camera images and the photodiode measurements. Each recorded image is associated with one pulse train of 10 consecutive laser pulses, while the successive trains are separated by $1\,\mathrm{s}$. Consequently, the 100-s acquisition sequence provides 100 independent realizations of the input signal, which are used for the statistical analysis of the shot-to-shot fluctuations.

\section{Spectrum evolution with power}
\label{app-spectra}

\begin{figure}[!htb]
\centering
\includegraphics[width=0.95\columnwidth]{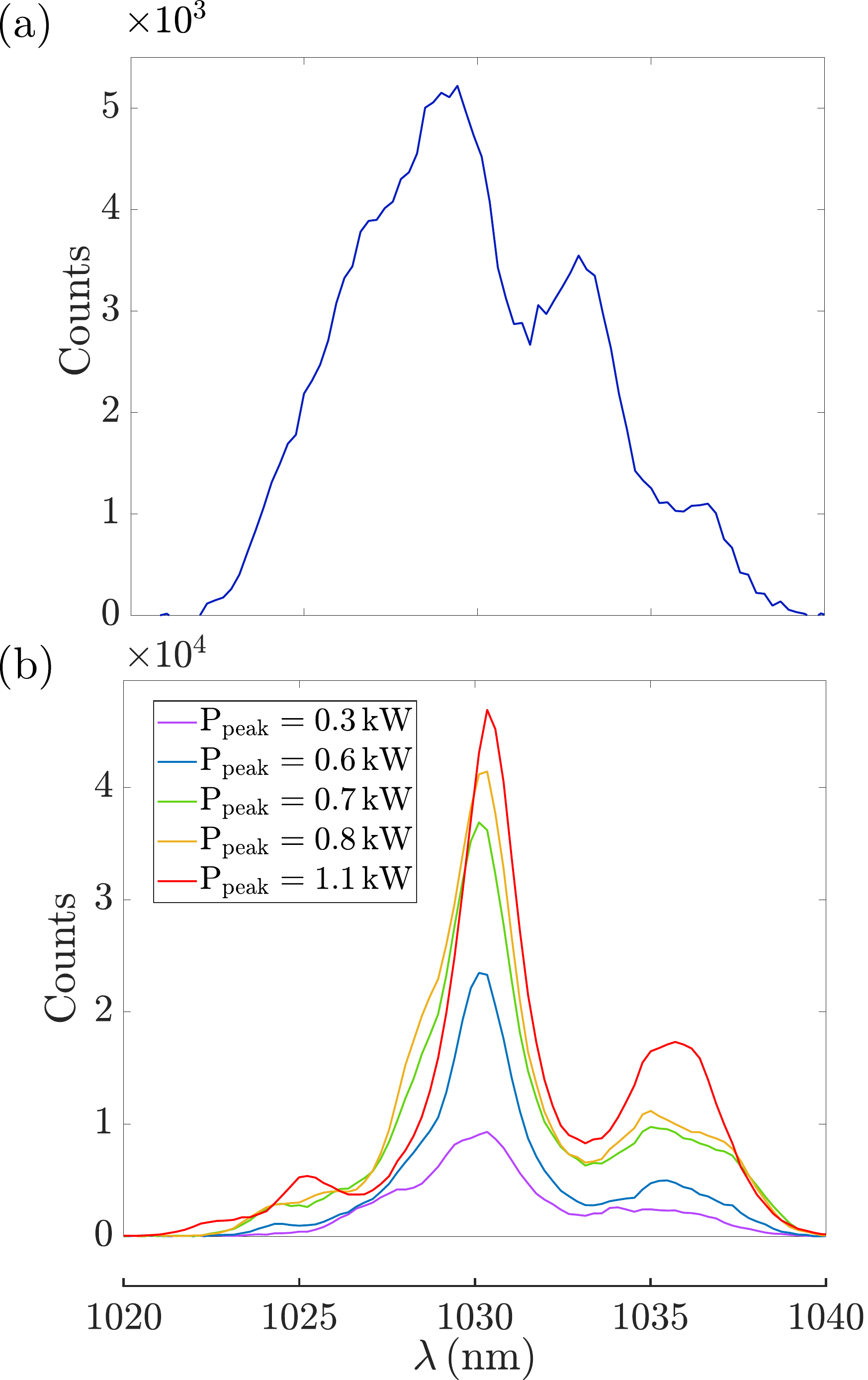}
\caption{Spectral characterization of the mode-locked laser source. (a) Input spectrum. (b) Output spectrum evolution with power.}
\label{fig:appC_spettri}
\end{figure}

In the case of the mode-locked laser, our spectrometer was capable of resolving the spectrum both at the fiber input and output. The former is characterized by two main peaks with an overall width of about 20 nm, as shown in Fig. \ref{fig:appC_spettri}a; the latter is reported in Fig. \ref{fig:appC_spettri}b at several values of the input peak power.

\section{Impact of noise}
\label{app-noise}

\begin{figure}[!htb]
\centering
\includegraphics[width=\columnwidth]{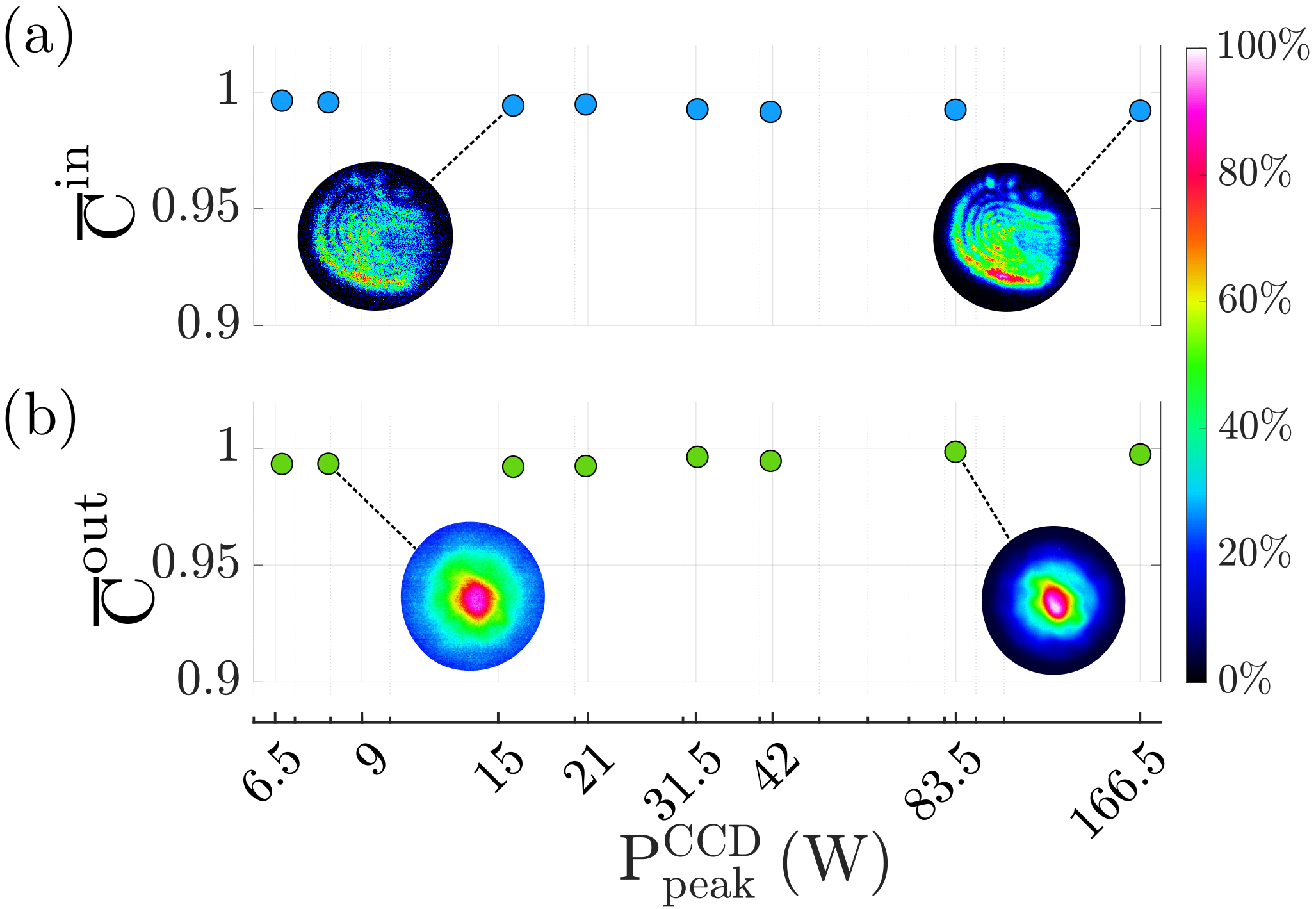}
\caption{Average input- and output-beam correlations as a function of the peak power incident on the CCD camera ($\mathrm{P_{\mathrm{peak}}^{\mathrm{CCD}}}$). Insets show representative beam profiles obtained with the Q-switched laser at the indicated incident powers. In particular, panel (a) shows the input beam, while panel (b) shows the output beam under fixed self-cleaning conditions. The power incident on the CCD is varied using attenuating filters placed before the camera, while the power launched into the fiber is kept constant to preserve the beam profile. In both cases, the average correlations remain close to unity throughout the investigated range, indicating a negligible influence of noise on our measurements.
} 
\label{fig:appB_noise}
\end{figure}

To rule out the possibility that the observed reduction of fluctuations with increasing power is merely a consequence of an improved signal-to-noise ratio at the camera, we performed a dedicated control experiment. The beam profile was kept unchanged, while its energy at the camera was varied by placing neutral-density filters in front of the CCD. We monitored the evolution of the correlation parameter ($\overline{C}$) as a function of the peak power reaching the camera, $P_{\mathrm{peak}}^{\mathrm{CCD}}$. This analysis was performed both for the input beam and for the output beam under self-cleaning conditions. In both cases, we found that $\overline{C}$ remains essentially unchanged over the investigated range of $P_{\mathrm{peak}}^{\mathrm{CCD}}$, despite the substantial variation in the detected power (see Fig. \ref{fig:appB_noise}). Therefore, the reduction of fluctuations observed with increasing power cannot be attributed to a simple improvement of the signal-to-noise ratio of the camera. Instead, these results indicate that the observed suppression of fluctuations is intrinsically associated with the nonlinear propagation dynamics.

\section{Gaussian distributions for a larger ensemble}
\label{app-1000acq}

\begin{figure}[!htb]
\centering
\includegraphics[width=\columnwidth]{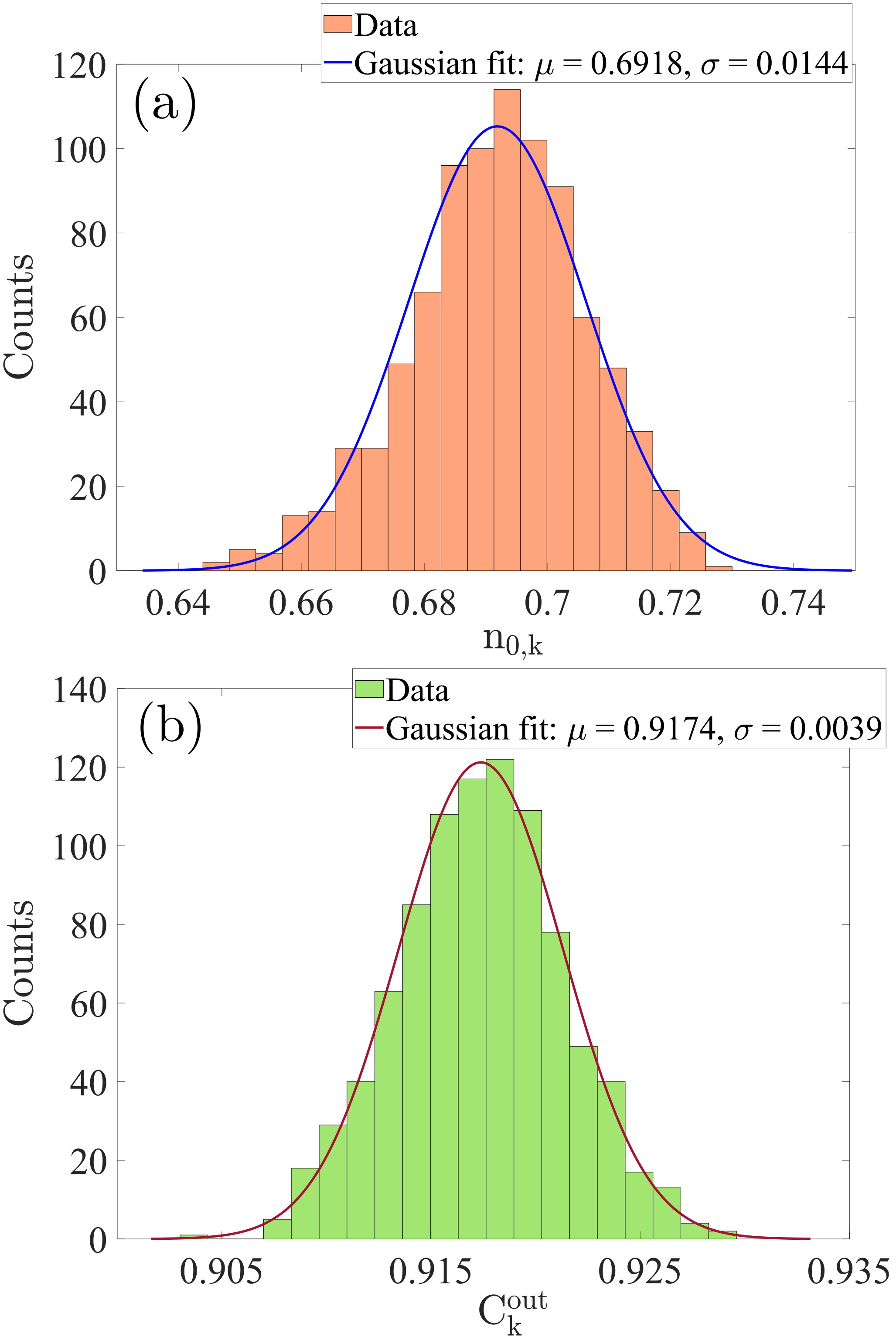}
\caption{Statistical distributions obtained from the enlarged ensemble of 900 realizations for the Q-switched source. (a) Distribution of the correlation with the fundamental mode at $P_{\mathrm{peak}} = 0.17$ kW. (b) Distribution of the correlation with the average image at $P_{\mathrm{peak}} = 7$ W. Solid lines show Gaussian fits to the experimental distributions. Here, $\mu$ and $\sigma$ denote the mean value and standard deviation of each distribution, respectively.}
\label{fig:appD_istogrammi}
\end{figure}

To assess the influence of finite statistical sampling on the observed fluctuation distributions, we repeated the measurements using a substantially larger statistical ensemble. In particular, 900 realizations were acquired with the Q-switched source. Figure \ref{fig:appD_istogrammi} shows the corresponding distributions of the correlations with the fundamental mode and with the average image, obtained at two representative input peak powers. In both cases, the distributions obtained from the larger ensemble are significantly closer to a Gaussian profile than those obtained from the standard 100-realization datasets. This indicates that the deviations from a Gaussian shape observed in the main measurements are at least partly associated with limited statistical sampling.

The larger ensemble, however, requires a substantially longer acquisition time and is therefore more sensitive to slow experimental drifts and variations of the experimental conditions. For this reason, the 100-realization datasets were retained for the main analysis, whereas the 900-realization measurements were used as an independent check of the underlying statistical distribution.

\end{document}